# Decarbonization of aviation via hydrogen propulsion: technology performance targets and energy system impacts

**Anna Cybulsky[1], Florian Allroggen[2,3], Yang Shao-Horn[4,5,6], Dharik S. Mallapragada[1*]**

1. MIT Energy Initiative, Massachusetts Institute of Technology, Cambridge, MA 02139
2. Laboratory for Aviation and the Environment, Department of Aeronautics and Astronautics, Massachusetts Institute of Technology, Cambridge, MA 02139
3. Joint Program on the Science and Policy of Global Change, Massachusetts Institute of Technology, Cambridge, Massachusetts 02139
4. Research Laboratory of Electronics, Massachusetts Institute of Technology, Cambridge, MA 02139
5. Department of Materials Science and Engineering, Massachusetts Institute of Technology, Cambridge, MA 02139
6. Department of Mechanical Engineering, Massachusetts Institute of Technology, Cambridge, MA 02139

* Corresponding author. Email: dharik@mit.edu

## Summary

The aviation sector is challenging to decarbonize since aircraft require high power and energy per unit of weight. Liquid hydrogen is an interesting solution due to its high gravimetric energy density, minimal warming impact, and low-carbon production potential. We quantify the performance targets for fuel cell systems and on-board storage to enable hydrogen-powered regional aviation. We then explore the energy infrastructure impacts of meeting this additional H2 demand in the European context under deep decarbonization scenarios. We find that minimal payload reduction would be needed for powering regional aviation up to 1000 nmi if fuel cell system specific power of 2 kW/kg and tank gravimetric index of 50% can be achieved. The energy systems analysis highlights the importance of utilizing multiple technology options: such as nuclear expansion and natural gas reforming with CCS for hydrogen production. Levelized cost of liquid hydrogen as low as €3.5/kg demonstrates pathways for Europe to achieve cost-competitive production.

## Context & scale

The aviation industry is under pressure to decarbonize and few viable solutions exist – the main ones considered are synthetic fuels (from biomass or power-to-liquid fuels from green hydrogen) and liquid hydrogen. Hydrogen use would require changes to aircraft designs and to the infrastructure network to supply the fuel. A hydrogen fuel cell system could be retrofit onto a regional turboprop aircraft with limited payload reduction, if such systems are optimized for high specific power. Using projected hydrogen aviation demand in a European case study, we show that the scale of sector-wide decarbonization required in 2040 implies ambitious expansion of clean power generation, regardless of aviation demand. Selected technology options will strongly influence system cost and the most economical allocation of generation and transmission between the hydrogen and power sectors. Central coordination between hydrogen producers and grid operators across national boundaries can help enable large scale deployment and increasingly affordable hydrogen cost.

## Introduction

Recent efforts to decarbonize the transportation sector, which contributed 37% of global end-use sector $CO_2$ emissions in 2021 [1], have largely focused on accelerating electrification of light duty vehicles [2] and developing commercial solutions for heavy-duty vehicles [3]. However, the shipping and aviation sectors are still in the early phases of developing new designs and technologies for decarbonization. Aviation faces perhaps the greatest hurdles to decarbonization due to the unique weight as well as volume constraints inherent to the application, which is compounded by the projected growth in aviation demand. By one estimate, the revenue-tonne-kilometers demand for aviation is projected to grow at 2.4-4.1% per year until 2050 [4], which if met via conventional aircraft, would lead to 1.4 to 2.3 times greater well-to-wake $CO_2$ emissions by 2050, as well as substantial non-$CO_2$ warming effects [4]. With advances in battery technology, there is growing commercial interest in battery-electric vertical take-off and landing (eVTOL) aircraft [5], as well as fixed wing commuter or regional aircraft – yet these segments represent a fraction of the total demand serviced by the industry [6]. Unlike battery-powered aviation, there is limited work on the performance requirements and energy system impacts of fuel-cell powered aviation, which is increasingly garnering more interest – for example, numerous start-ups are working on different types of fuel cell powered aviation strategies and the European manufacturer, Airbus, is designing a hydrogen aircraft for commercialization in 2035 [7]. In this context, this article seeks to quantify the aircraft technology performance targets and potential energy system impacts of converting the regional and short-range aviation segment (up to 1000 nautical miles (nmi)) from combustion to hydrogen powered fuel cell-based electric powertrains.

Although direct electrification of end-uses is desirable to minimize energy losses, this may be particularly challenging for aviation due to limitations on battery specific energy density. For example, Viswanathan et al [8] have estimated specific energy requirements at 1,200-2,000 Wh/kg at the pack level for aircraft, which is roughly 4-6 times that of state-of-art Li-ion battery packs used in light duty vehicle transportation, while suggesting that gravimetric energy levels up to 600 Wh/kg could be achieved by the end of this decade [8]. As compared to electrification, use of synthetic fuels, either produced from biomass or using electricity and biogenic carbon feedstocks, has been the primary focus of the industry to date, given their compatibility with existing aircrafts. However, a combination of growth in aviation demand and competition for biomass availability by other end-uses might limit the scalability of this approach for sectoral decarbonization [9, 10, 11]. The reliance on emerging power-to-liquid fuels technologies could also supply additional liquid fuel but is likely to have greater power sector investment needs as compared to hydrogen use due to lower efficiency of electricity and energy use when considering the complete fuel production supply chain [12].

These challenges with battery electrification and synthetic fuel routes have led to increased interest in hydrogen use for aviation, as part of a portfolio approach for sectoral decarbonization [4]. As compared to jet fuel and battery electrification, hydrogen is appealing for aviation not only because it does not produce $CO_2$ emissions at the point of use but also because of its larger gravimetric energy density (e.g. 120 MJ/kg vs. 44 MJ/kg for jet fuel). Use of hydrogen instead of synthetic fuel also avoids further energy efficiency losses associated with fuel production. Despite this, the use of hydrogen for aviation faces a number of challenges relating to aircraft design as well as cost-effective sourcing of the quantity and form of hydrogen and associated energy system implications. **First**, even the most volumetrically dense form of hydrogen, liquid hydrogen, has over 4x lower volumetric density compared to jet fuel (8 MJ/L vs. 35 MJ/L). **Second**, the gravimetric energy density advantage of using hydrogen vs. jet fuel is partly offset when accounting for the weight of the insulated tank that is used to store the liquid hydrogen at -253°C. **Third**, although hydrogen fuel cells avoid NOx emissions and reduce contrail and cirrus effects compared to combustion engines [13], the relatively low specific power of fuel cells vs. gas turbines (12-23 kW/kg[14]) remains a challenge (see later discussion). **Fourth**, the use of low-carbon hydrogen at scale requires its cost-effective production, as well as scale up of the hydrogen transmission and storage infrastructure and accompanying grid infrastructure [15, 16, 17]. All of these factors need further analysis at both aircraft and system-level to evaluate the viability of hydrogen-powered aviation as part of global decarbonization efforts.

While prior work has mainly evaluated the use of fuel cells for commuter aircraft and hydrogen combustion for the short-range segment [13, 18, 19], the regional segment could be a strong candidate for introducing fuel cells in aviation due to the lower power requirements as compared with narrow-body aircraft [13], and the potential to serve a significant share of flights by achieving ranges up to 1000 nmi [6]. The introduction of these aircraft will depend not only on the viability of aircraft design, but also on the infrastructure and supply chain network required. Here, we address both these questions to quantify the potential for hydrogen-powered regional aviation. Our key contributions include: a) developing a simplified weight-constrained model for evaluating performance targets for fuel cell systems and H2 storage for retrofitting regional aircraft for H2 propulsion, b) quantifying hydrogen demand for regional aviation using a performance model [20] and flight data for the case study in western Europe and c) evaluating the bulk electricity and hydrogen infrastructure requirements and energy system impacts of meeting the above hydrogen demand for aviation under deep decarbonization constraints.

We apply the weight-constrained model to the regional Dash 8 turboprop aircraft, and find that it could be retrofit with limited to no payload reduction (up to 8%) if a tank gravimetric index of 35-50% and a fuel cell system specific power of 2 kW/kg are achieved. These findings are broadly aligned with other more detailed assessments on system design for regional aviation [13, 18, 21] and highlight the appeal of the simplified approach. Using a plausible aircraft technology configuration and operating scheme, we estimate that powering regional aviation for a flight network spanning 171 airports across five countries in Western Europe will consume 4.3 Mtonnes H2, which is approximately 29% of the projected hydrogen demand for other end-uses by 2040. What are the energy system impacts of serving this additional hydrogen demand? We explore this question using a 76-node energy system model of the region where we co-optimize bulk electricity and hydrogen infrastructure under deep decarbonization constraints [22]. Across various technology scenarios, we find that increased H2 demand from aviation increases power generation capacity, H2 generation capacity and annual system cost by 5%, 36% and 12%, respectively with the base case assumptions. Unsurprisingly, pathways with lowest cost rely on investment in the broadest technology options for H2 and power infrastructure, including pipelines and liquid trucks for H2 transport, electrolyzers, $CO_2$ capture and storage (CCS) based technologies for H2 and power generation, as well as nuclear power. In the absence of CCS technologies, H2 production tends to be spatially distributed, due to primary reliance on electrolyzers as well as decreased investment in H2 pipelines, while CCS deployments leads to more centralized H2 supply chains. Liquid truck deployment allows for locating liquefiers in locations with cheap low-carbon electricity and enables more flexibility in the development of the H2 supply chain. Overall, the energy systems analysis highlights pathways to achieve large-scale H2 production in coordination with the power sector, with system-wide levelized cost of liquid hydrogen falling between 3.53-4.68 €/kg.

The next section describes the results on the performance requirements for fuel cell and H2 storage systems, which is followed by results on aviation H2 demand estimation and associated energy system impacts. This is followed by a discussion characterizing the implications of the results and identifying areas for further work. A concise description of the methodology is provided at the end with further details available in the supporting information (SI) and accompanying master's thesis [23].

# Results and Discussion

## H2 propulsion and storage systems for regional aircraft retrofit

To evaluate key technology metrics for a hydrogen-powered aircraft, we focus on regional aircraft for a retrofit analysis, characterized here by the turbo-prop De-Havilland Dash 8-400 aircraft, which can accommodate 78 passengers. This single aircraft was responsible for 3.96 Mtonnes of $CO_2$ emissions in 2019 (or 0.5% of global emissions), while the regional aircraft segment[1] represents 7% of global $CO_2$ emissions, 4% of revenue passenger kilometers (RPKs) and 29% of departures [6]. Although this segment represents a minority of emissions and RPKs, it is a good entry point for aviation electrification, as smaller aircraft with shorter range, lower maximum take-off weight (MTOW)[2], and lower efficiency make better candidates for early electrification [8, 24, 25], as constrained by current battery specific energy or fuel cell specific power. They also represent the highest $CO_2$ intensity-segment in terms of $gCO_2$/RPK [6]. Finally, retrofits constitute a lower-risk certification strategy for short-term commercialization, since only those systems affected by the change in powertrain may need to be re-certified.

In the retrofit analysis, we make two inter-dependent assumptions to simplify the analysis about the technology performance targets for hydrogen fuel aircrafts that circumvent the need to consider details of the aerodynamics of aircraft design. First, we assume that the new fuel (or energy storage medium) and propulsion system would be constrained to match the absolute weight allocated to the current systems, while the fundamental aircraft design is held constant – such as fuselage and wing dimensions. Thus, the current weight of the maximum fuel carried and the engines provides a weight capacity constraint for the hydrogen powertrain system of 6,800 kg. The new system, consisting of electric motors, fuel cells, hydrogen tank, and possibly a reduction in payload, would have to fit in within this weight envelope to keep overall aircraft weight constant (see Equation 1 in Experimental Procedures). Second, because we are keeping the aircraft total weight constant, the power requirements are also implicitly constrained to be the same as the conventional aircraft, set based on the maximum rated power of the engines of 7562 kW, as needed for take-off [26].

[1] The regional segment typically seats between 50-120 passengers [6].

[2] MTOW includes operational empty aircraft weight, payload (passengers, crew, cargo), and fuel weight (including reserves).

With these two assumptions, there remain two key variables for determining range and/or necessary payload reduction to enable hydrogen-powered regional aviation: a) fuel cell to propulsive power efficiency, which determines the amount of $H_2$ fuel to be stored on board for a given range and b) fuel cell specific electrical power density (in kW/kg) that determines the weight of the propulsion system to meet the power requirements. These two fuel cell parameters are inter-related by the operating current density of the fuel cell. For example, one can operate the fuel cell at a low current density point on the polarization curve to achieve high efficiency, with the obvious benefit of carrying less hydrogen fuel onboard and decreasing weight allocated to $H_2$ storage. At the same time, operating at low current density implies an increase in stack area and potentially weight to meet the necessary stack-power-rating constraint.

The different power requirements during various phases of the trip (take-off, climb, cruise, descent) imply that the fuel cell efficiency (and operating current density) will also be different for each phase. The maximum power requirements are determined by take-off and climb phases, which are met by operating proton exchange membrane (PEM) fuel cells at the lowest energy efficiency (typically at about 50%). Consequently, due to the inverse relationship between efficiency and current density, the fuel cell will need to be operated at higher efficiency (lower current density) during cruise and even higher in descent, when power requirements are lower. For our baseline calculations of fuel requirement for the trip, we used: a) a trip-average PEM fuel cell efficiency of 60% and b) baseline fuel cell system specific power of 1 kW/kg, including the balance of plant systems (thermal management/cooling, air management including compression), which is defined based on the maximum fuel cell power. According to the U.S. Department of Energy, a stack-only specific power of 2 kW/kg has been achieved, while only 650 W/kg has been achieved at the system level in automotive applications – although these values are dated [27] and higher values of 2.9 kW/kg have been cited since [18]. In addition, we assumed a liquid hydrogen tank gravimetric index (GI), which represents the ratio of fuel weight to full tank weight, of 35%, similar to what was assumed for the short-range aircraft in [13]. High indices of 83-84% have been achieved in large-scale space applications [19], and 25% has been achieved in shuttle on-board $LH_2$ storage of 100kg [19]. On the other hand, automotive applications thus far have focused on compressed gas tanks with low tank GI of 6% [19].

Given these baseline assumptions, the retrofit aircraft would require 522 kg of hydrogen with a full tank weight of 1500 kg for the lowest range of 500 nmi. Figure 1A highlights that retrofitting such a configuration would require a 34% payload reduction. Moreover, the largest weight increase in the hydrogen fuel-cell powered aircraft compared to the conventional aircraft comes from the powertrain system, highlighting the importance of design improvements in this area.

While increasing tank gravimetric index will generally allow for heavier payload capacity, the impacts are greatest at longer ranges when fuel requirements are greater. For example, Figure 1B shows alternative designs with zero payload reduction, which highlights that a tank gravimetric index (GI) of 50% would require a fuel cell (FC) specific power of 1.9 kW/kg at a range of 1000 nmi, while a GI of 35% would require FC specific power of 2.4 kW/kg for the same range. It is also worth noting that a combination of improvements, such as a tank gravimetric index of 50% and a fuel cell specific power of 1.5 kW/kg requires no payload reduction up to about 600 nmi and requires just 13% payload reduction at 1000 nmi. To maintain the same passenger traffic, a 13% reduction in aircraft payload could imply approximately 13% more flights.

Figure 1C illustrates that higher fuel cell specific power strongly impacts payload reduction. For example, doubling specific power from 1 kW/kg to 2 kW/kg decreases payload reduction from 58% to 8% for a range of 1000 nmi. Within the next decade, a re-design of system components (including both stack and balance of plant) with optimization for minimal weight could enable higher power density, such as through the use of composite bipolar plates.

**Figure 1: Hydrogen aircraft retrofit results for different assumptions on fuel cell efficiency, specific power (SP) as well as H2 storage tank gravimetric index (GI). A) Comparing weight distribution of conventional and proposed fuel cell powered regional aircraft for 1000 nmi under various technology assumptions B) Specific power required to achieve zero payload reduction with variable gravimetric index and assumed average fuel cell efficiency of 60% C) Payload reduction required (in % of aircraft payload at a given range) to meet variable target fuel cell specific power under gravimetric index constraint of 35% and average fuel cell efficiency of 60%. Negative values imply that payload capacity would be larger than the baseline.**

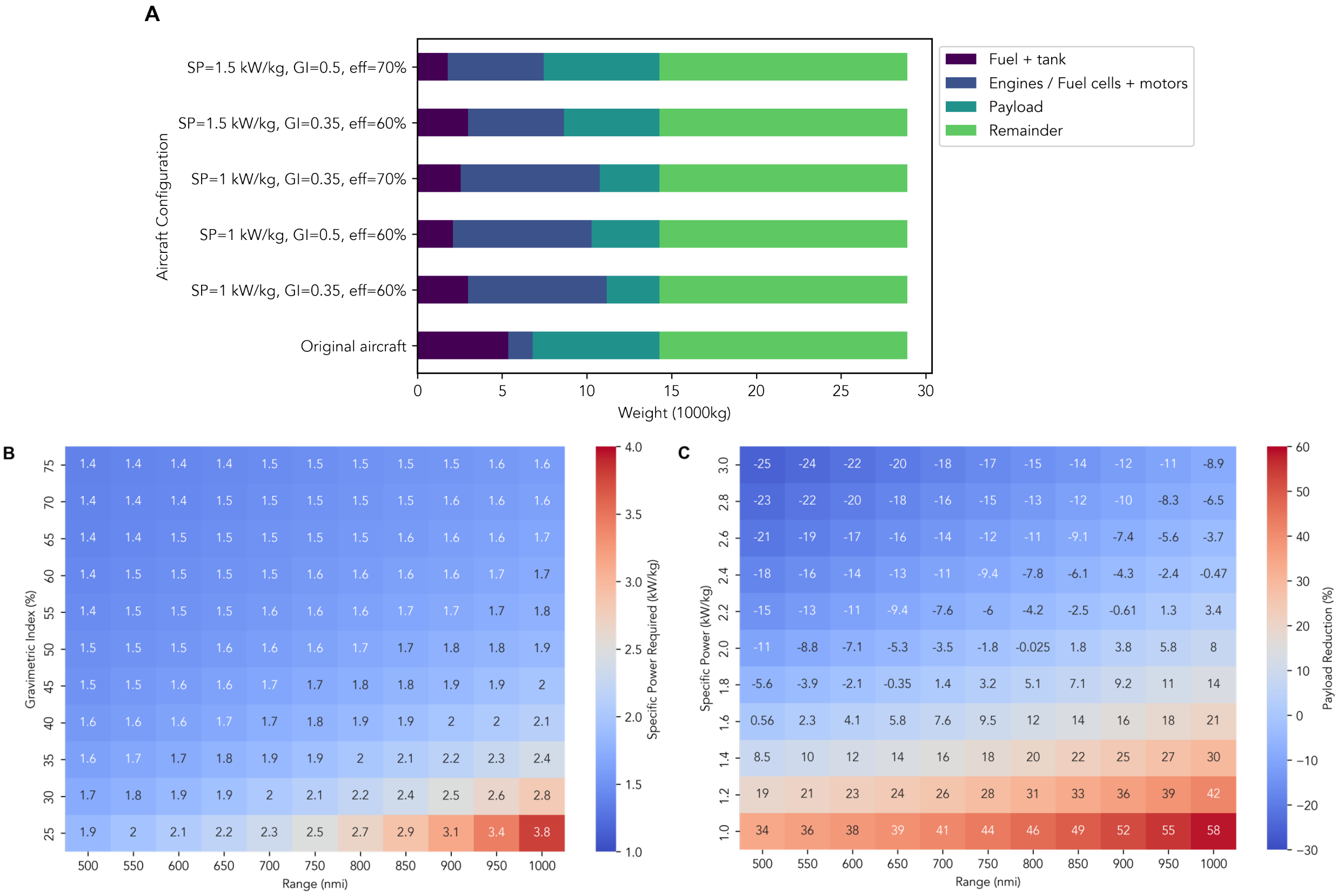

For the given aircraft, with 1000 nmi range, Figure 1A shows increasing fuel cell efficiency via operating at lower current densities does not have as big of an impact on payload as changing tank GI or FC specific power. This is because operating at lower current densities leads to increased stack size to meet the power requirement constraint that more than offsets weight savings in fuel storage for the baseline assumptions. However, operating at high fuel cell efficiency is accompanied with lower heat losses, resulting in smaller cooling systems, which could compensate for some of the increase in stack weight and still allow for more payload at higher efficiencies– this factor is not accounted for in our analysis. Meanwhile, an improvement in fuel cell polarization curve, which effectively enables higher voltage at the same current density, will be beneficial and can achieved by reducing kinetic, transport, or ohmic losses.

It's useful to contextualize our findings versus the literature. Other studies on hydrogen fuel-cell propulsion use on aircraft have assumed that fuel cell specific power between 1.6-4 kW/kg would be required, with a range between 30% to 60% tank gravimetric index [13, 18, 21]. The higher end of the fuel cell specific power requirement considered cooling and air compression requirements separately [18], implying that the system-wide specific power requirement would be effectively lower, reaching values similar to this study. It was also considered that for aircraft carrying greater passenger numbers for the short-range segment, fuel cell systems could be used in hybridization with combustion H2 engines [13], while hybridization with batteries was considered in [21], where it was found that significant improvement to battery specific energy would be required to be beneficial.

## Hydrogen aviation demand in the context of a European case study in 2040

We evaluate the impact of hydrogen demand from aviation and associated energy system impacts using the case study of regional aviation across five countries in Western Europe: France, Germany, Italy, Spain and the United Kingdom. These countries are among those with the largest number of airports and the largest aviation demand for the regional aircraft segment (up to 1000 nmi) in Western Europe. As hydrogen aircraft with this payload capability and range are unlikely to reach the market before 2035, the year 2040 is used as a baseline for the analysis. The aviation demand model estimates a total of 0.5 EJ of annual jet fuel consumption based on 2019 flight data, which translates into 4.3 Mtonnes of annual hydrogen fuel required across the region, with a range of 0.71 to 0.95 Mtonnes per country (see Figure 2, liquid demand). For the energy systems analysis, we consider aviation H2 demand in 2040 along with projected demand for hydrogen from other sectors (Figure 2, gaseous demand) [28, 29], including industry and road transportation. With these assumptions, aviation accounts for 28.6% of total hydrogen demand, and tends to be more concentrated than demand from other sectors. Electrical demand projections for 2040, taken from [28, 30], point to increases of 20% to 55% by 2040 vs. 2016 (6.8 EJ to 9.1 EJ total, see demand map in Figure S1), resulting from increasing electrification of sectors such as transportation and building end-uses.

**Figure 2: Projected aviation hydrogen demand (liquid) and base hydrogen demand (gaseous, from other sectors) in 2040. Data sourced from OAG flight schedule data (2019) on a daily time scale, which is distributed evenly across all hours of the day, and includes 171 airports, for which demand was allocated to the closest node.**

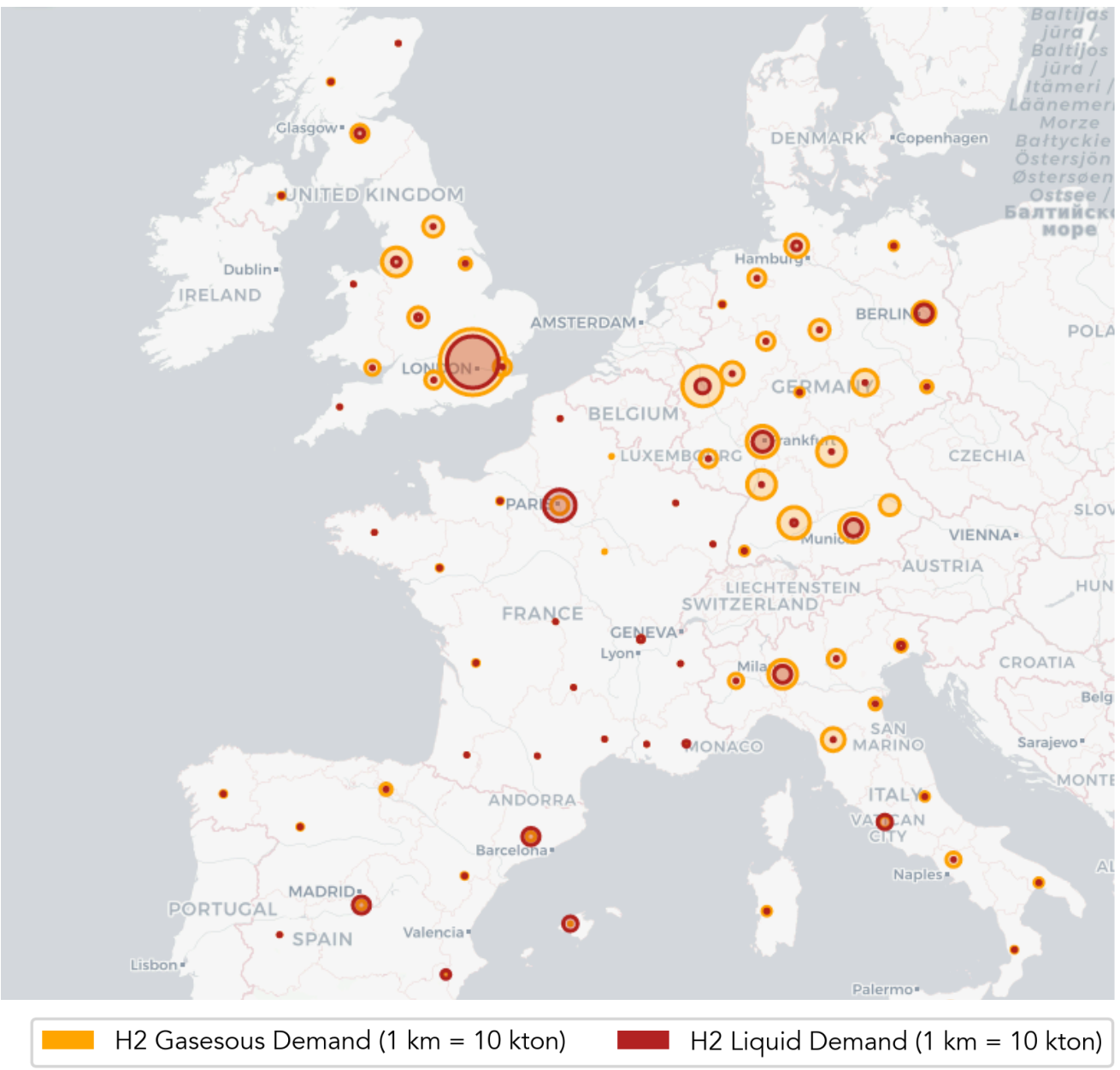

## Energy system impacts of aviation H2 demand

We use the open-source energy systems model, Decision Optimization of Low-carbon Power and Hydrogen Networks (DOLPHYN – see further details in methods and section SI [31]) to evaluate the least-cost investment and operation of bulk electricity and hydrogen infrastructure to meet the specified electricity and hydrogen demand under various technology and policy scenarios, summarized in Table 1. We consider a 76-node representation of the region of interest, as shown in Figure 2, while considering 50 representative days of system operation (at an hourly time scale) to approximate annual operating outcomes. Across the two supply chains, we consider a range of technology options for generation, transmission, and storage as summarized in SI. All scenarios include a combined sectoral $CO_2$ emissions constraint of 38 Mtonnes of $CO_2$, which counts all combustion related emissions modeled and corresponds to 86% reduction vs. 2018 levels for the power sector (while effectively the economy-wide reduction is greater since other sectors are being decarbonized through electricity and hydrogen via increasing electrical loads).

**Table 1: Scenarios analyzed via the DOLPHYN model for the 76-node region for Western Europe analyzed here.**

| Scenario Name | Description |
|---|---|
| Base | Hydrogen demand in other sectors (including industrial, other transportation; excluding building heating). Nuclear expansion is not allowed, CCS is allowed in certain locations. Electrical transmission expansion is allowed (up to 10 GW) and hydrogen pipeline expansion is allowed (no limit). |
| Base + Aviation | All assumptions as Base case, with additional liquified hydrogen demand for aviation. |
| With Nuclear Expansion | All assumption as Base + Aviation case, except that nuclear expansion is allowed in places with current nuclear capacity (based on 2017). |
| No Carbon Capture | All assumptions as Base + Aviation case, except that CCS is not allowed |
| No CCS with Nuclear Exp. | All assumptions as Base + Aviation case except that CCS is not allowed, while nuclear expansion is allowed, in places where it currently exists (based on 2017). |
| Liquid Trucking | All assumptions as Base + Aviation case, with additional option of using Liquid trucking on routes up to 500 km. This implies that liquefaction may not need to occur onsite, at every airport. |
| No Pipelines | All assumptions as Base + Aviation case except that hydrogen pipeline construction is not allowed. |

## Hydrogen use for aviation impacts hydrogen infrastructure more than electricity infrastructure

Figure 3 highlights that adding hydrogen demand for aviation leads to more dramatic impacts on hydrogen infrastructure compared to the electricity infrastructure, in part because not all of hydrogen supply is sourced from electrolysis (left-most bar in Figure S2C). In particular, 42% of the additional baseload $H_2$ demand from aviation is met by natural gas autothermal reforming (ATR) with carbon capture and sequestration (CCS). As compared to the Base case, an additional 1000 tonnes/hour or 36% greater capacity is required in the Base + Aviation scenario (Figure 3C), even though total $H_2$ demand increases by 40% - slightly increasing overall electrolyzer capacity factor by 2%. Additional storage is deployed in the aviation case – on the order of 100 GWh or 3000 tonnes of combined liquid and gaseous storage (Figure 3E). This amount of storage is sufficient to meet average aviation demand for a period of about 4 hours. The storage installations help balance hourly demand with supply in a cost-effective manner by reducing electricity consumption for liquefaction and electrolytic hydrogen production during high electricity prices and increasing production during low price periods. The share of liquid $H_2$ storage (34% of $H_2$ storage in the Base + Aviation case) also enables relatively high capacity utilization of the liquefier capacity (79%) that is sized to be a factor of 1.3X (about 21 $GW_{H2}$ or 625 tonnes/hour) of the average liquid hourly liquid $H_2$ demand from aviation.

The partial reliance on $CO_2$-emitting ATR w/ CCS technology for meeting additional $H_2$ demand means that greater emissions reductions are necessary in the power sector for Base + Aviation demand scenario vs. the Base scenario. Consequently, we see increased VRE capacity and decreased natural gas capacity in the Base + Aviation scenario vs. the Base scenario, while total capacity increases by about 5% (55 GW) (Figure 3A). Considering the sizable effects of exogeneous electrical load increase combined with emissions restrictions on the power sector in 2040, production of hydrogen for aviation will not be the main driver of power sector expansion.

**Figure 3: Comparing different energy system scenarios referenced to the Base + Aviation case. All scenarios consider Base + Aviation hydrogen demand, except S0 – Base demand only. A) Electrical capacity B) Electrical generation C) Hydrogen capacity D) Hydrogen production E) Storage capacity F) Transmission capacity**

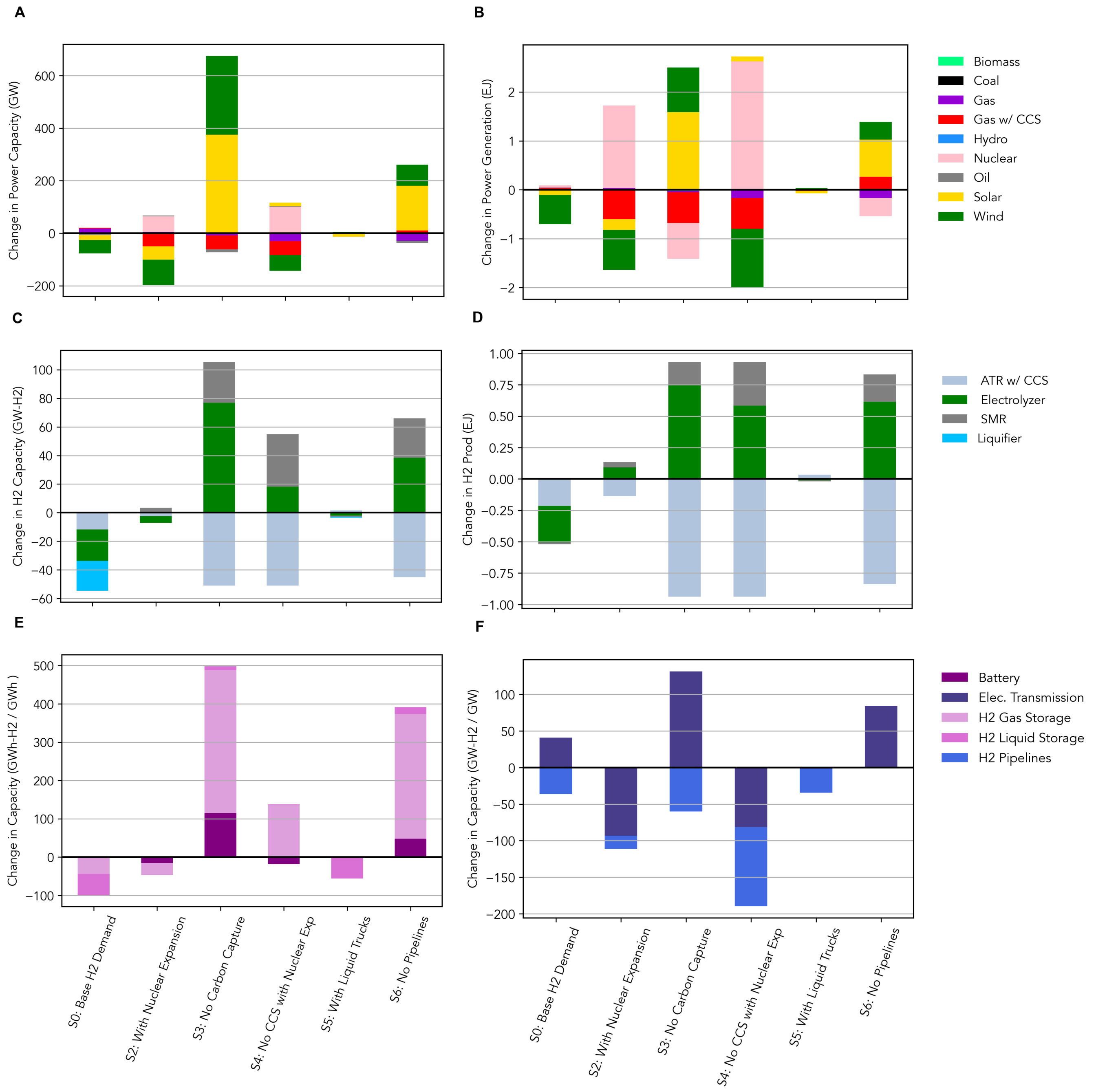

## Utilizing all possible technology options is crucial to accelerate decarbonization

In terms of technology options, the effects of allowing nuclear expansion or not deploying CCS technologies produce significant changes in the results. For example, without CCS technologies, over 600 GW of additional VRE capacity is required, representing greater than 50% increase compared with a total of 1130 GW required in the Base + Aviation scenario (Figure 3A). Aside from the impact on the electrical sector, where no gas power plants with carbon capture can be deployed, the no carbon capture scenario implies producing significantly more hydrogen via electrolysis, which adds over 300 TWh of additional electricity demand onto the system (equivalent to 10% of baseline load) (Figure 3B, 3C). Furthermore, this scenario gives the largest hydrogen storage and electrical transmission requirements – an additional 370 GWh (11 $H_2$ ktonnes) and 130 GW respectively compared to the Base + Aviation scenario (see Figure 3E, 3F), due to the predominant reliance on VRE generation. This would imply total new transmission lines equivalent to 72% of current installed capacity.

Irrespective of CCS technology availability, allowing the expansion of new nuclear reduces reliance on VRE capacity and gas power generation with CCS, while increasing deployment of natural gas reforming via SMR for H2 production (Figure 3A and 3C). The latter happens because nuclear expansion makes it cost-effective to further reduce power sector emissions than the Base + Aviation scenario which creates additional emissions budget for use of fossil-based H2 production. The lowest capacity system is achieved in the nuclear expansion scenario (where nuclear power is only allowed to expand in the zones where it currently exists, based on 2017 data). Interestingly, the baseload nature of nuclear power generation implies that electrolyzer capacity utilization increases with nuclear expansion as well as total production via electrolysis, while electrolyzer capacity deployment reduces (see Figure S7 for sample dispatch). The scenario with nuclear *and* no CCS demonstrates that nuclear expansion is able to replace the large scale VRE expansion seen in the No CCS scenario. Moreover, as this scenario leads to largely distributed hydrogen infrastructure, hydrogen storage replaces pipelines (see Figure 3E vs 3F) – albeit system costs are still about 5% greater than in the Base + Aviation case.

## Carbon capture creates a decentralized network whereas electrolysis has a distributed nature

Figure 4A highlights the spatial trends in the H2 infrastructure in the Base+Aviation case, where ATR resources are concentrated in locations with carbon storage around the North Sea, and electrolysis in areas with high VRE resource quality (e.g. coastal UK, coastal Spain, Italy). This is indicative of a decentralized hydrogen network i.e. a network displaying characteristics of both centralized and distributed networks. The network is more centralized around the regions with CCS in France and Germany, using pipelines to distribute large quantities of hydrogen, and then is more distributed in the UK, Italy, and Spain. France becomes a hydrogen exporter, producing a total of 4.8 Mtonnes of hydrogen.

In the scenario without CCS and therefore increased electrolysis, the system becomes more distributed, with two major generation centers in southern UK (which has the largest wind capacity) and a large pipeline leading to Germany (Figure 4B). The UK, with the highest share of VRE generation among the five countries, becomes the largest hydrogen producer with 3.8 Mtonnes of production, while France, with the largest share of nuclear, sees its production reduced to 2.7 Mtonnes. The most distributed $H_2$ network occurs in the scenario with no carbon capture *and* nuclear expansion (see maps in Figure S4). As nuclear is allowed to expand in all countries (except Italy, which has no current capacity) along with transmission lines, average wholesale electricity prices become more level across all zones (Italy excluded, see Figure S5). Therefore, since ATR with CCS is no longer a technology option, electrolysis becomes more distributed according to demand levels. Moreover, this scenario requires the lowest pipeline capacity – over 7x less than in the Base + Aviation case, as well as 80 GW less of electrical transmission capacity (see Figure 3F).

**Figure 4: Results from the energy system model showing illustrative maps of the hydrogen and electricity systems. A) Hydrogen system for Base + Aviation scenario B) Hydrogen system for No CCS scenario C) Electricity system for Base + Aviation scenario D) Electricity system for No CCS scenario. The size of the bubble is proportional to the indicator shown (i.e. hydrogen demand or generation, electricity demand or generation, and transmission capacity)**

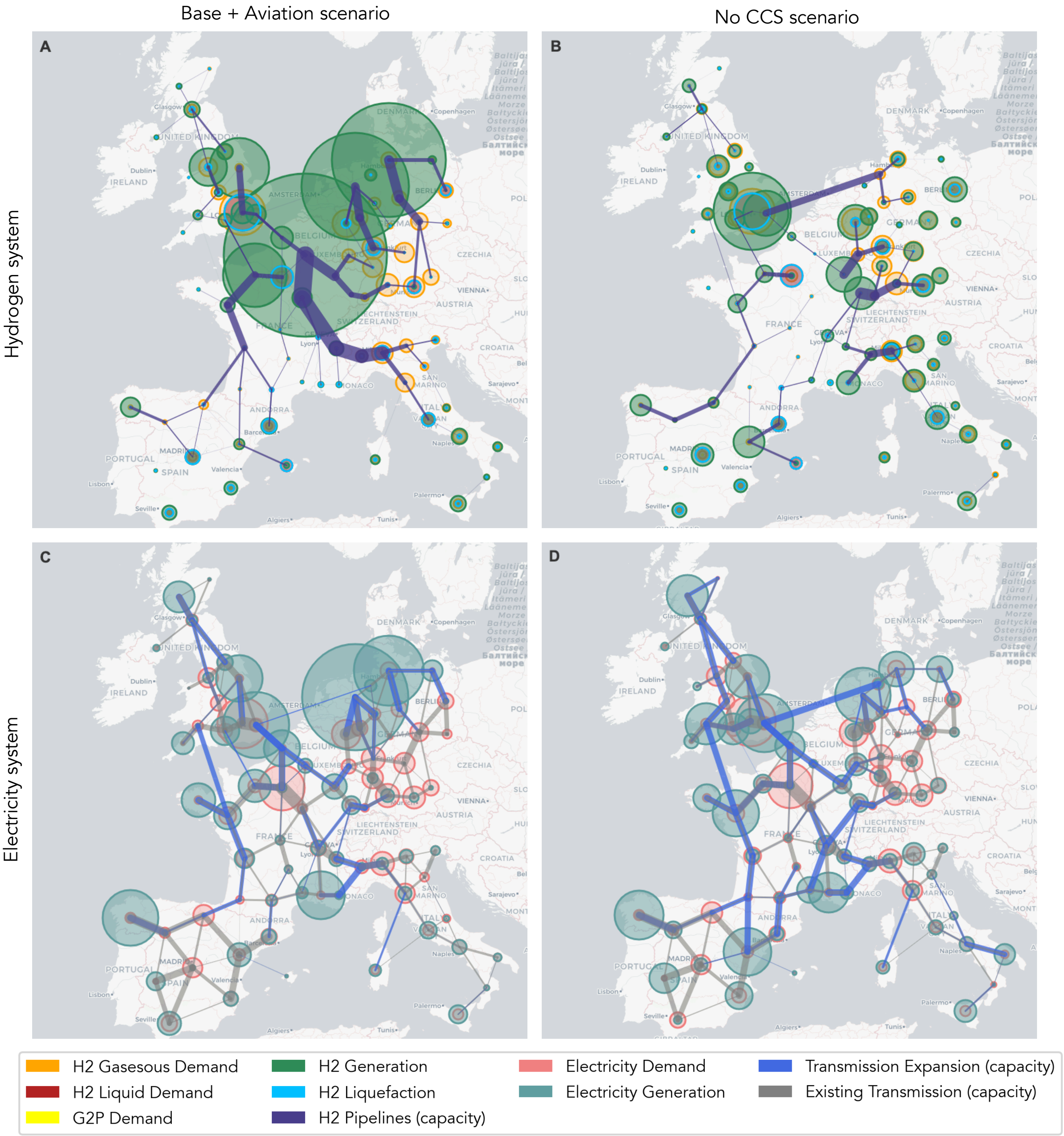

**Figure 5: Electrolytic hydrogen production, liquefaction, demand and electricity price per zone A) Base + Aviation scenario, showing how electrolysis location compares to total demand B) With Liquid Trucking scenario, showing how liquefaction location shifts with respect to liquid H2 demand**

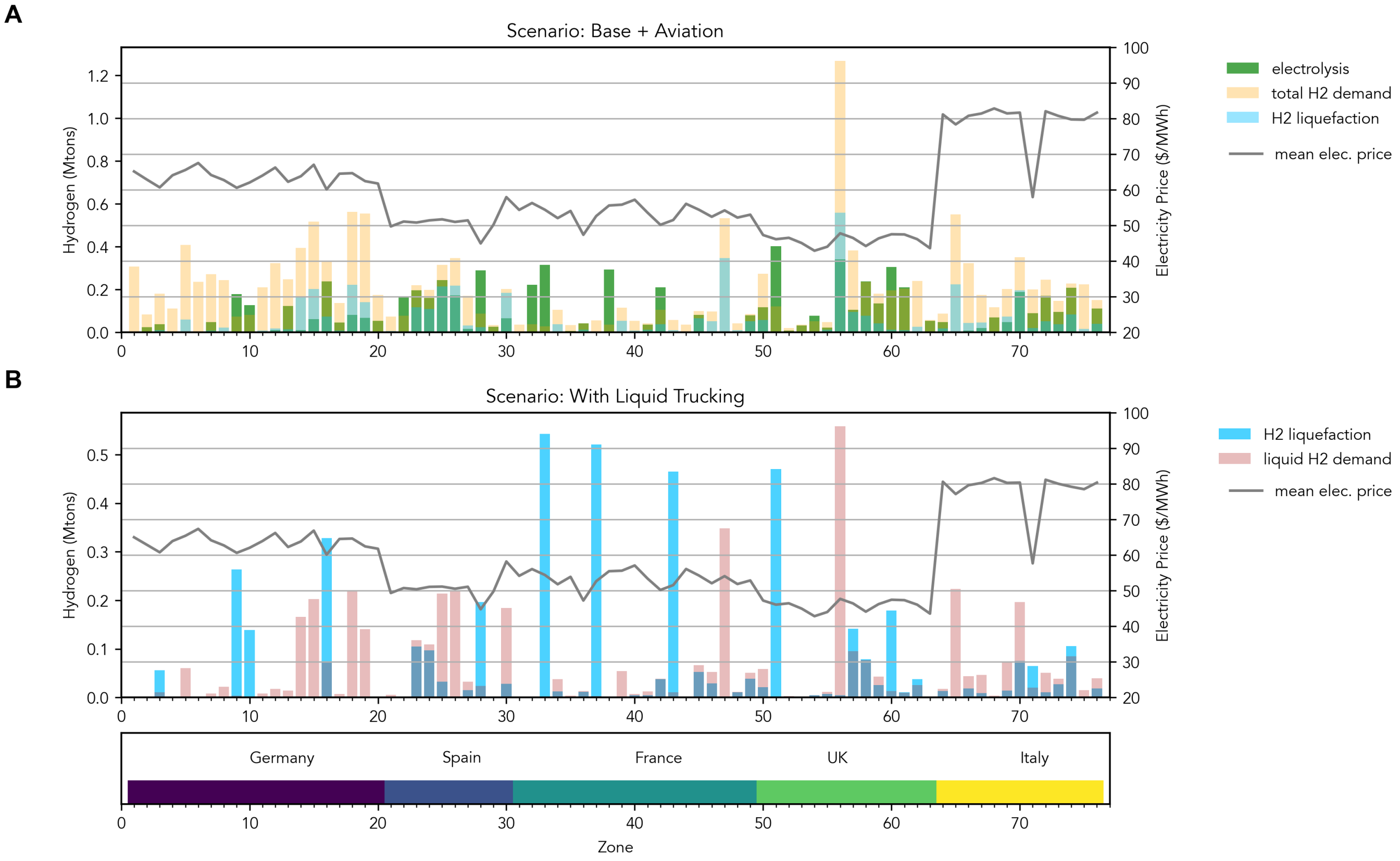

## Liquid H2 transport via trucks enables simultaneous storage and transport

The liquid trucking scenario allowed to assess whether trucks could lower system costs, compared to the exclusive use of gaseous H2 pipelines. Without trucking, hydrogen must be liquified at the point of demand – and thus, at each airport. The results show that liquid trucking is indeed deployed with 330 trucks for the region, allowing to shift liquefaction capacity to areas with lower electricity prices (see Figure 5B), and results in similar system cost compared to Base + Aviation case (see Figure S6). Interestingly, while pipeline investment is reduced, no liquid $H_2$ storage is built, implying that trucks are used both as a storage and transmission means (Figure 3E & 3F). This suggests that although trucking is not vital for large scale hydrogen transmission, they could be deployed, and may allow for gradual development of the H2 infrastructure to support aviation. In a few zones, the number of trucks deployed by the optimization model is not realistic - for example, charging or discharging over 240 trucks per day at a given location. Therefore, trucking will be most viable for airports that are within a few hundred kilometers to other zones with higher renewables availability or lower electricity prices, but only for airports that have a lower level of demand up to a couple hundred tonnes per day – leading to about 50 truck deliveries per day.

### The scale of expansion required is ambitious but is economically feasible

The system cost of the Base + Aviation case is 12% greater than the Base case (see Figure S6), highlighting the incremental bulk energy system cost of meeting the additional H2 demand for 1000 nmi aviation segment (which equates to an average incremental cost of €3.5/kg$H_{2liq}$). However, the Base + Aviation case was modeled with the same emissions cap as the Base case, and thus includes the cost of additional economy-wide decarbonization for the aviation sector, which may need to occur via other means if not via hydrogen. As a consequence, the marginal carbon abatement cost rises from €230 per tonne for power sector decarbonization (including increasing electrification, but without any hydrogen demand), to €310 per tonne for cross-sector decarbonization using hydrogen. New (absolute) power capacity required is estimated at 760 GW in the Base + Aviation scenario, considering 140 GW of retired capacity (mostly coal, OCGT, and CCGT – although CCGT capacity remains). Considering existing capacity, this represents 5% more total capacity than in the Base case. In addition, 330 GW (or 76 TWkm) of transmission expansion would be required, equivalent to 50% of current capacity (or 56% in TWkm). Considering a time frame of 15 years, this translates into an average annual VRE capacity deployment rate of 50 GW/year. For hydrogen, achieving the required 130 $GW_{H2}$ of capacity would require the equivalent of 9 $GW_{H2}$ per year of construction, with about a quarter to a third dedicated to aviation. Annual system costs are estimated by the model at €145 B in the Base + Aviation scenario, equivalent to just over 1% of GDP of the five countries studied [32]. Therefore, although the scale of capacity expansion is ambitious, it does appear to be plausible.

In the Base + Aviation scenario, about 20% of the systems emissions cap is allocated to the hydrogen sector, or 7.7 Mtonnes of $CO_2$. However, this does not include emissions from electricity production used for electrolysis. Estimating these emissions using average power emissions intensity for each country and annual electrolyzer generation implies an additional 3.84 Mtonnes of $CO_2$ emissions. Given that aviation demand represents 28.5% of total $H_2$ demand, approximately 3.3 Mtonnes of total $CO_2$ would stem from $H_2$ production, while no $CO_2$ emissions would be produced during flight. For comparison, the jet fuel burn in conventional aircraft for the given fight network is estimated to produce 32.2 Mtonnes of $CO_2$ in flight, based on jet fuel emissions coefficients [33]. Therefore, $CO_2$ emissions would be reduced by around 90%, depending on the electricity and hydrogen technology mix.

If the same aviation segment was decarbonized using synthetic fuels created via the Fischer-Tropsh pathway [12], the required quantity of input hydrogen would be at least 1.7 times greater. Well-to-wheel emissions reductions for transportation are estimated to be in a similar range of 90-108%, as emissions from vehicle operations are offset by biogenic $CO_2$ embedded in the synthetic fuel (lower reduction range applies to plants with $H_2$ recycle vs. without $H_2$ recycle) [12]. Therefore, the direct use of hydrogen makes for a very strong argument for reducing total amount of input hydrogen and consequently bulk energy system infrastructure investment needs, while synthetic fuels would require more limited end-use infrastructure investment.

Finally, the model demonstrates that at large scale, levelized cost of hydrogen can be relatively low, at a system-wide average below €2/kg for gaseous hydrogen and as low as €3.53-€3.66/kg for liquid hydrogen in three of the scenarios (see Figure S6), but rising up to a maximum of €4.68/kg. This cost does not include last-mile distribution and refueling costs, which are estimated to increase LCOH between €0.13/kg to €0.27/kg [13]. Assuming a hydrogen fuel cost of €4/kg, this is equivalent to €0.12/kWh and approximately €1.2/L for jet fuel. Given that hydrogen fuel cells or hydrogen combustion engines would achieve similar (or higher) efficiencies to jet fuel engines, hydrogen fuel costs could be cost-competitive with jet fuel on a per propulsive energy basis in 2040, under the evaluated scenarios.

## Discussion

In this work, we first identified fuel cell and hydrogen storage performance targets for regional aviation with minimal payload reduction (10% or less). We find that targeting an increase in fuel cell system specific power to 2 kW/kg will be most impactful to reduce propulsion system weight, followed by an increase of tank gravimetric index to 50%. The combination of these two would lead to a retrofit aircraft with unchanged payload capacity, as per our simplified weight-based model. Next, we estimated aviation demand for five countries and a flight network up to 1000 nmi in Europe and find that annual total hydrogen demand would be in the order of 4.3 Mtonnes or 28% of total projected hydrogen demand for 2040. Finally, we evaluated implications of scaled up use of hydrogen aircraft in the context of broader expansion of the grid and hydrogen infrastructure for economy-wide decarbonization. We find that incremental hydrogen demand from aviation is met by a portfolio of technologies (electrolysis, ATR w/ CCS, liquid storage) and results in power generation capacity, H2 generation capacity and system cost increasing by 5%, 36%, and 12% respectively compared to the base case. Expanding the portfolio of technologies reduces the power system capacity requirement and overall system costs. For example, the expansion of nuclear power reduces power capacity by over 10% and annual system costs by around 3% (see Figure S6). The evolution of the hydrogen network ranges between decentralized (e.g. in the Base + Aviation scenario) to distributed (e.g. No CCS scenario) depending on the portfolio of generation technologies. Trucks can reduce the number of pipelines and average pipeline size required, but will be limited to serving smaller airports for practical reasons. Finally, the overall cost of hydrogen can be relatively low when considering future cost projections that depend on scaling of infrastructure, and could be on par with jet fuel prices at €1.2/L.

While the modeling is intended to assess relative trade-offs between different policies and technologies rather than absolute planning scenarios, the results should also be considered in terms of the practicality of their implementation. Carbon capture may play a critical role in creating large clusters of natural-gas based hydrogen production, such as the one seen in northern France. However, in practice, this region vastly lacks sufficient carbon storage projects (considering those announced thus far [35]) to produce such a large quantity of hydrogen via

ATR with CCS. On the other hand, Germany and the UK have announced much larger carbon storage projects [36, 37], which could accommodate the additional carbon – in practice this likely means relocating ATR plants to those regions, which would impact the network. Furthermore, the amount of nuclear expansion observed in the scenario where it is allowed is equivalent to 67% expansion or 60 GW of capacity over 15 years. Given that constructing these power plants can take 5-10 years, their design and safety assessments would need to begin by 2030. In the No CCS with Nuclear scenario, expansion reaches 100 GW or over 100%, which may not be realistic by 2040.

For hydrogen aviation deployment, locations that exhibit local production and cheaper electricity can make good candidates for first adoption of hydrogen aviation infrastructure. For instance, in the Base + Aviation scenario, Hamburg airport (in northern Germany) is well positioned, as it could locally develop hydrogen generation capacity needed to meet about 80% of total demand, and its average electricity price is slightly cheaper than in neighboring areas thanks to renewables availability. In Spain, where baseline hydrogen demand is lowest, there are multiple candidates: Barcelona, and the regions along the southern coast (such as Alicante, Seville and Malaga). In France, the southern region near Marseille and the northern region near Nantes are well suited to produce hydrogen to meet local demand. In the UK, the northern region around Glasgow and Edinburgh could locally meet their hydrogen demand, if transmission lines were expanded from up north where wind availability is higher. Heathrow airport could generate hydrogen as well, up to about 30% of its total demand, and would require new electrical and/or hydrogen transmission from the northern zone near Hull. Finally, in Italy, where average electricity prices are highest, only the Sicilian zone comes close to meeting its demand, while Rome is able to meet about half of its overall demand. Taranto and Naples are also reasonable candidates. All of these zones, with the exception of Italy, are able to produce liquid hydrogen below €4/kg (in the Base+Aviation scenario) with the lowest LCOH found in the UK. An initial hydrogen flight network could be established between these candidates. Generally, all these locations have good renewables availability and/or transmission capacity to nearby generation sites.

The analysis further demonstrates pathways for producing cost-competitive hydrogen in Europe. In May 2022, the EU Commission published the REPowerEU plan, which plans for 10 Mtonnes of hydrogen production in 2030, as well as 10 Mtonnes of imports [38]. At an average system-wide LCOH below €2/kg for gaseous H2 in three scenarios, estimated in this study using technology cost assumptions for 2040, Europe could in theory become self-sufficient in hydrogen production. However, the pace of infrastructure buildout required could leave room for hydrogen imports, if domestic production cannot be scaled quickly enough due to various factors. While the modeling assumed a central planner viewpoint with perfect coordination, the reality will be different. Nonetheless, this study highlights the value of coordination between the hydrogen and power sectors.

In order to meet target emissions limits to achieve net zero by 2050, this study demonstrates that planners must consider all possible technology options. The development of carbon capture and storage and/or nuclear power generation are key enablers for sector-wide decarbonization. Policies that support the expansion of these technologies will allow to reduce system costs by lowering the required capacity of VRE generation required and allowing to achieve improved capacity utilization across both sectors. While new nuclear capacity has higher capital costs than solar or wind [39], this analysis considers full system costs including transmission and storage costs – both for electricity and hydrogen, which may be disregarded in simpler technology-level analyses. Moreover, lack of carbon storage limits the use of natural gas reforming for hydrogen production and increases dependence on electrolyzers – which in turn implies greater reliance on electricity generation. While carbon storage may not be perceived as viable in many regions, due to its continued reliance on fossil fuels, its potential to accelerate decarbonization in the medium-term should not be neglected.

While H2 demand for aviation and its energy system impacts have not been studied in the literature, our results on H2 and electricity system infrastructure are broadly aligned with other studies on sector-coupled energy system expansion under carbon constraints. As an example, studies using the PyPSA-Eur model have highlighted the importance of sector coupling (across heating, transport, and electricity) in lowering energy system costs as well as reducing the need for transmission expansion [17, 40]. Similar to our study, many analyses have highlighted the substitution between energy storage and transmission as well as between pipelines for H2 transport and electricity transmission [17, 40]. That said, our study's characterization of truck vs. pipeline transport as well as considering impacts of CCS use in power and H2 sectors, and doing so with high spatial resolution are differentiating features compared to the existing literature.

Future work could seek to further refine the assessment of H2 use for aviation and its energy systems impacts in several ways. First, the analysis can be extended to consider longer range aircrafts and competition with other net-zero solutions (synthetic fuels, biofuels, direct air capture), which were not considered here. Second, the energy system modeling can be refined to account for economies of scale associated with key H2 infrastructure technologies like liquefaction, pipelines, and ATR with CCS. In addition, allowing for retrofit of existing natural gas pipeline networks may represent a more realistic baseline for pipeline expansion. Expanding the network to include neighboring countries, such as Belgium, Denmark, and the Netherlands, could provide a more comprehensive picture of Europe's potential for meeting its future electricity and hydrogen needs and the impact of demand for $H_2$ from new segments like aviation. In terms of emissions assessment, we did not account for $H_2$ fugitive emissions which will have non-zero indirect radiative forcing effects in the medium-term [55], as well as upstream emissions in the fossil fuel supply chain (e.g. for natural-gas based hydrogen production). A more refined assessment could be done to evaluate such upstream and downstream emissions more holistically – along with the cross-sector implications on emission reduction targets when the decarbonization of multiple sectors occurs simultaneously.

# Experimental Procedures

*Lead contact: Dharik Mallapragada, dharik@mit.edu*
*Materials availability: https://github.com/anna-nadia/DOLPHYN_results*
*Data and code availability: https://github.com/macroenergy/DOLPHYN, https://github.com/anna-nadia/DOLPHYN_scenarios*

## Hydrogen Aircraft Retrofit Model

To provide a rough baseline for key technology metrics for electrification of aircraft via hydrogen, we select the regional turbo-prop aircraft De-Havilland Dash 8-400, which can accommodate 78 passengers, for a retrofit analysis. The underlying assumption of the retrofit model is that the overall weight of the aircraft must remain the same, and therefore any hydrogen systems must fit within the weight envelope of the former conventional systems (Eq. 1). To estimate the amount of hydrogen fuel required for a given range, we first estimate the actual energy required for the full aircraft mission of 1100 nmi, including reserves, while considering the differences in energy efficiency of the conventional engine and fuel cells.

**Eq. 1 – Weight balance constraint for retrofit analysis**

$$\begin{aligned} &Maximum\ useable\ jet\ fuel\ weight + Engines\ weight \\ &\qquad = Electric\ motors\ weight + Fuel\ cell\ weight + Hydrogen\ tank\ weight \\ &\qquad + Payload\ reduction \end{aligned}$$

The weight of the electric motors for the specified power requirement is estimated based on high-temperature, super-conducting electric motors, which have achieved specific power of over 12 kW/kg [43]. For comparison, the original aircraft engines on the DHC-8-400 have a specific power of 5 kW/kg – allowing the difference to be allocated to the new system.

For the best in-class turboprop engine thermodynamic efficiency of 35% [14] and the maximum onboard fuel energy stored of 65,600 kWh, the actual energy required for the full aircraft mission of 1100 nmi (including reserves) is estimated to be 23,000 kWh. To consider the impact of aircraft range, we assume that the energy requirement is proportional to the range and remains constant at about 21 kWh/nmi (for this aircraft), which is generally a reasonable assumption between a minimum range of 500nmi and up to the maximum aircraft range. This relationship has been validated for this range using the Aviation Emissions Inventory Code (AEIC) performance model [20].[3] The same relationship can be applied to the retrofit aircraft, since the structure and thus the aircraft aerodynamics are unchanged (lift-to-drag ratio), the same propellers convert shaft power to propulsive force giving a constant propulsive efficiency, and the overall weight (MTOW) remains the same. For a conventional aircraft, flying shorter distances implies carrying less fuel onboard, potentially resulting in lower initial aircraft mass and therefore a lower average mass throughout the flight. In the hydrogen retrofit analysis, the new aircraft utilizes the full weight capacity at all ranges. Using the Breguet-Range equation [43], which relates aircraft range to initial and final aircraft mass (with the difference being fuel burned), it is verified that the impact of lower initial aircraft weight on fuel burn at lower ranges (up to 500 nmi) can be neglected.

[3] This approximation was verified with data from the AEIC model, but the ratio begins to deviate (increase) below 500 nmi.

Below we provide the equations used for the retrofit model calculations.

To calculate the energy required for a given flight range, we use Eq. 2.

**Eq. 2**

$$energy_{required} = \frac{desired\ range}{maximum\ aircraft\ range} * energy_{\mathrm{full_{tank}}} * thermal\ efficiency\ engines$$

Next, the amount of energy is converted to a hydrogen quantity using the lower heating value of hydrogen and the average assumed fuel cell efficiency (60%) in our calculations, using Eq. 3.

**Eq. 3**

$$hydrogen_{weight} = \frac{energy_{required}}{LHV\ \left(\frac{kWh}{kg}\right) * fuel\ cell\ efficiency_{average}}$$

The hydrogen weight can be converted to a weight for the full hydrogen tank, based on the tank gravimetric index – one of the main input assumptions that we vary in our calculations, using Eq. 4.

**Eq. 4**

$$hydrogen_{tank_{weight}(full)} = \frac{hydrogen_{weight}}{tank\ gravimetric\ index}$$

Finally, we solve for the output parameter desired (either fuel cell specific power) or payload reduction, using a combination of Eq. 5 and Eq. 1.

**Eq. 5**

$$fuel\ cell\ weight = \frac{engines\ rated\ power}{fuel\ cell\ specific\ power}$$

## Aviation Demand Model

In order to approximate aviation fuel demand for a given flight network in Europe, current flight and fuel burn data is required. The AEIC model [20] is a detailed performance model that is used to calculate aircraft fuel consumption for the flights of interest. Flight schedule data from OAG for 2019 is filtered on the following criteria, as a feasible segment for initial conversion to hydrogen aircraft:

- Flights of less than 1000 nmi

- Aircraft with up to 220 seats (to eliminate wide-body aircraft)
- Flights departing from the selected countries in Europe: France, Germany, Italy, Spain, United Kingdom

Jet fuel demand is converted to hydrogen fuel demand by taking into account the relative energy densities of the two fuels, and adding 10% to account for the possibility of heavier aircraft [13, 19]. Possible differences in engine efficiency are not accounted for due to the wide range of aircraft that may be replaced (e.g. turbofan engines with higher efficiencies than turboprop engines) and the possibility that hydrogen aircraft design may utilize combustion engines rather than or in addition to fuel cells. Demand is aggregated on a daily basis, and is spread throughout the day evenly. This produces an annual hourly load profile. The hydrogen demand for each airport is spatially allocated to the closest node modeled up to a distance of 231km, while airports that are farther from a given node are considered to be out of geographic scope of the model. 75% of airports were allocated to a node within a distance of less than 107 km.

In addition to aviation, other sectors are projected to consume hydrogen as well, such as road transport (heavy duty and light duty vehicles), shipping, as well as industry – referred to as hydrogen base load in this article. Demand projections for these sectors for each country are taken from the TYNDP Scenario's Report of the ENTSO-E and ENTSO-G [28], with the exception of the UK [29]. Heating demand was excluded since there is wide debate about hydrogen's suitability for this purpose [44] and the sizable demand would have significant impact on the scale of the hydrogen system. As hydrogen load profiles are not available, base demand is distributed spatially per zone according to electrical loads (which are approximated based on GDP and population [45]), while temporal demand is distributed evenly with 45% in summer months and 55% in winter months. Since hydrogen infrastructure is interdependent with the electricity sector, for example through electrolysis technology, electrical demand growth must also be accounted for. Electricity demand projections are taken from the same source [28] for the year 2040, with the exception of the UK which used an average load increase of 31%, in alignment with [30]. Total demand for the five countries grows from 1898 TWh to 2514 TWh. Electrical load profiles are based on load profiles from 2016 data compiled by PyPSA-Eur [45].

## Energy Systems Model

In order to model the development of the hydrogen sector in conjunction with the power sector for the year 2040, the open-source Decision Optimization of Low-carbon Power and Hydrogen Networks (DOLPHYN) energy system optimization model is used [15, 16, 31]. The model can be parameterized with cost and technical characteristics (e.g. efficiency) for production, storage, and distribution technologies for electricity and hydrogen. The region of interest (five countries in Europe) is described as a spatial network comprising 76 nodes. The temporal resolution is at the hourly scale, thus 8760 hours in a full year can be considered. To reduce the size of the problem, a time-domain reduction algorithm reduces all time-dependent data to 50 representative days (with hourly resolution), using a k-means clustering algorithm.

The objective function of the model minimizes overall system costs: annualized capital costs for technology expansion (production, transmission, and storage), fixed operational costs, as well as fuel costs and other variable costs associated to existing and new capacity. Costs are calculated and summed for each zone and each time step. The model also makes decisions on which capacity should be retired.

A number of constraints are applied on the model. The major constraint in the model is the power balance, which ensures that power or hydrogen supplied at each time step and zone must equal the level of demand at each time step and zone. Supply can be provided through production, storage discharge, or through transmission arriving at a given zone, and conversely will be reduced through charging storage, departing transmission (exports), or losses (power line losses or boil-off losses for hydrogen). Figure 6 illustrates the balance for hydrogen – both gaseous and liquid.

**Figure 6: Overview of power and hydrogen balance in the DOLPHYN energy systems model**

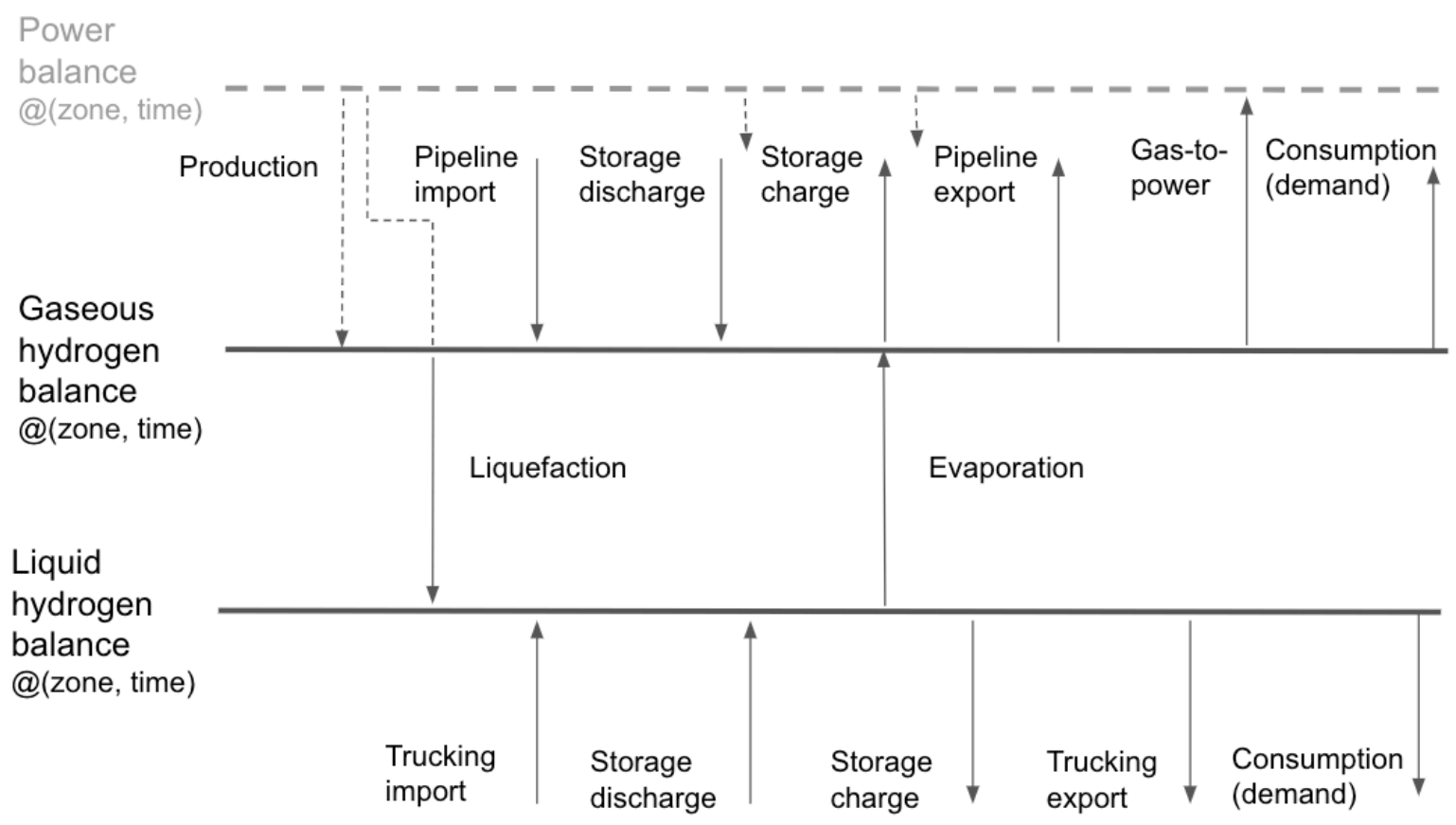

Another significant constraint applied on the model is the emissions cap, which is applied jointly to the electricity and hydrogen sectors, or as one shared global constraint. For the year 2040, power sector emissions are reduced by 89% compared to 2018 levels, as suggested by the TYNDP Scenario's Report [28]. Given that electricity demand growth stems in part from the electrification of other sectors (e.g. light-duty vehicles), the cap effectively has a larger impact on economy-wide emissions as opposed to power sector decarbonization only .

The electricity and hydrogen sectors are coupled through three main pathways: gas-to-power, conditioning (compression, liquefaction), and power-to-gas. Power-to-gas refers to electrolysis technology, through which electricity is used in the production of hydrogen from water. Gas-to-power refers to the use of hydrogen in fuel cells or combustion of hydrogen in CCGT plants for power generation. Finally, compression of hydrogen gas, which is generally required for pipelines and for storage, as well as liquefaction of hydrogen both consume electricity - which is added as demand to the electricity sector.

Seven scenarios are developed to explore the effects of aviation demand and different technology mixes for the power and hydrogen sectors. (see Table 1) The Base scenario assessed the system with baseline hydrogen load only. Aviation demand was then added on top, before changing other factors. In the baseline scenarios, nuclear power capacity was not allowed to expand, and technologies requiring carbon capture and storage technologies could only be deployed in regions around the North Sea where storage projects have been announced [35, 36, 37]. Maximum transmission expansion was set to 10 GW per line, including for lines that have been announced but are not yet operational. Pipelines are allowed to expand without limit, except in scenario 6. Trucks are not available except for scenario 5, where their potential is assessed.

An average cost of hydrogen for the system is calculated by taking the total hydrogen sector costs, including variable electricity costs and pipeline network, and dividing by the total amount of hydrogen production. The variable costs of electricity are accounted for by using the marginal electricity prices (calculated as the duals of the optimization problem, which represent the cost of producing one additional unit of electricity at a given time) at the time of hydrogen generation. The costs of liquefiers are accounted for separately, as well as variable electricity costs for liquefaction, and divided by the total amount of hydrogen liquified for aviation – which are then added on top of the gaseous hydrogen production and transmission costs. See Figure S6 for results and Equation S1.

## Acknowledgements

The authors would like to acknowledge Youssef Shaker's contributions in gathering input data for the energy systems analysis and processing the results from the model. We would also like to acknowledge the support of Dr. Martha Frysztacki from the PyPSA group in helping to prepare input data for the model. D.S.M acknowledges funding from the MIT Energy Initiative Future Energy Systems Center. A.C. and Y.S.H. acknowledge funding from the MIT Energy Initiative Low-Carbon Energy Centers for Energy Storage.

## Declaration of Interests

Anna Cybulsky is currently an employee of ESMIA Consultants. The authors declare no competing interests.

## Supplemental Information

Technology cost assumptions are listed in the tables below. A discount rate of 4% is used to annualize and report the present value of technology costs for the model, which uses yearly costs as inputs. This rate is based on the real (post-inflation) WACC used by the EIA Annual Energy Outlook, and is similar to the real WACC rate described in the EPA IPM model [46].

1. Electrical generation resources considered for expansion in the model are:

- CCGT and OCGT (open-cycle gas turbines)
- CCGT with CCS
- Solar
- Wind (onshore and offshore)

2. Resources not considered for expansion (only existing resources are modeled):

- Biomass
- Coal
- Hydro (including run-of-river)
- Nuclear (except in scenarios 6 and 7)
- Oil
- Geothermal

### Calculation of average LCOH

**Equation S1: Calculation of system-wide average LCOH for gaseous and liquid hydrogen. Note: pipeline costs include compressor costs. Trucking fixed and operating expenses are included for scenario 5.**

$$\begin{aligned} LCOH_{gaseous} = {} & H2\ production\ fixed\ costs + H2\ storage\ fixed\ costs \\ & + H2\ pipeline\ fixed\ costs + H2\ variable\ electricity\ costs \\ & + fuel\ costs\ (natural\ gas) / \ Total\ H2\ generated \end{aligned}$$

$$\begin{aligned} LCOH_{liquid} = {} & H2\ liquefaction\ fixed\ costs \\ & + variable\ electricity\ costs\ /\ Total\ H2\ liquified \end{aligned}$$

**Table S1: Power sector technology cost assumptions. Heat rates are based on data of existing plants (except for nuclear, where data was unavailable), and the average per country is used (here the overall average is shown).**

| **Technology** | **Lifetime[4]** | **Capital Cost** | **FOM** | **VOM** | **Heat rate** |
|---|---|---|---|---|---|
| **Name** | **Years** | **EUR/kW** | **EUR/MW-yr** | **EUR/MWh** | **MMBTU/MWh** |
| Onwind [47] | 30 | 1040 | 25486 | 2.3 | N/A |
| Offwind [47] | 30 | 1930 | 44484 | 2.7 | N/A |
| Solar [47] | 25 | 600 | 25000 | 0.01 | N/A |

[4] Lifetime is used to annualize costs. For data from [48], lifetime refers to capital recovery period.

| | | | | | |
|---|---|---|---|---|---|
| PHS [47] | 80 | 2000 | 20000 | 0 | N/A |
| Hydro [47] | 80 | 2000 | 20000 | 0 | N/A |
| Run of river [47] | 80 | 3000 | 60000 | 0 | N/A |
| OCGT [47] | 30 | 400 | 15000 | 3 | 9.75 |
| Nuclear [47] | 30 | 4776 | 128794 | 8 | 10.44 |
| CCGT [47] | 30 | 800 | 20000 | 2.65 | 6.56 [56] |
| Coal [47] | 40 | 1300 | 25000 | 6 | 10.04 [56] |
| Lignite [47] | 40 | 1500 | 30000 | 7 | 10.04 [56] |
| Geothermal [47] | 40 | 3392 | 80000 | 0 | 14.28 [56] |
| Biomass [47] | 30 | 2209 | 100000 | 0 | 11.77 [56] |
| Oil [47] | 30 | 400 | 6000 | 0 | 10.34 [56] |
| CCGT- 90% CCS [48] | 20 | 1392 | 52929 | 5.3 | 7.16 |
| Battery [47] | 15 | 167kW-1 / 122KWh-1 | 29993 | 0 | 0.85 RTE[5] |

**Table S2: Fuel costs [47]. The cost of transporting and storing carbon is included in the cost of fuel for plants with CCS, using a cost of 30 EUR/ton $CO_2$ in Europe [49].**

| Fuel | EUR/MMBTU |
|---|---|
| natural gas | 6.33 |
| uranium | 0.88 |
| coal | 2.46 |
| lignite | 0.85 |
| biomass | 2.05 |
| oil | 14.65 |
| natural gas with CCS – 90% capture | 7.76 |
| Natural gas with CCS – 94% capture | 7.82 |

**Table S3: Transmission assumptions, costs are based on the PyPSA-Eur model [45]**

| Transmission lines | Reinforcement Cost (annualized) [26] EUR/MW-km-yr | Line losses [50] %/100-km |
|---|---|---|
| HVAC lines | 47.5 | 0.625 |
| HVDC lines (existing, planned or under construction) | 14.3 - 284 | 0.625 |

**Table S4: Hydrogen production technology assumptions**

[5] RTE: round trip efficiency

| Technology | Lifetime | Investment Cost | FOM | Efficiency |
|---|---|---|---|---|
| Name | Years | EUR/kWH2 | EUR/kWH2/year | % LHV |
| PEM Electrolyser [51] | 40 (10 for stack) | 375.4 | 18.5 | 65 |
| ATR w/ 94% CCS [52] | 25 | 851.7 | 25.5 | 67.5 |
| SMR [53] | 25 | 840.2 | 39.5 | 76 |
| Liquifier with Terminal [54] | 40 (30 for terminal) | 2732.7 | 98.6 | 9 kWh/kg |

**Table S5: Hydrogen gas-to-power technology assumptions**

| Technology | Capital Recovery Period | Investment Cost | FOM | VOM | Efficiency |
|---|---|---|---|---|---|
| Name | Years | EUR/kW | EUR/MW/year | EUR/MWh | % HHV |
| CCGT-H2 [48] | 20 | 702 | 24700 | 1.76 | 54 |
| OCGT-H2 [48] | 20 | 594 | 18525 | 4.41 | 35 |

**Table S6: Hydrogen storage technology assumptions**

| Technology | Lifetime | Investment Cost | Compressor Cost | FOM |
|---|---|---|---|---|
| Name | Years | EUR/ton | M-EUR/ton/hour | EUR/ton/year |
| Gas tanks (underground) [55] | 30 (15 for compressors) | 494,276 | 1.47 | 1011 |
| Liquid/cryo tanks [54]* | 30 | 133,507 | | 9436 |

*boil-off losses are considered

**Table S7: Hydrogen pipeline technology assumptions**

| Technology | Lifetime | Investment Cost | Compressor Cost | Boosters |
|---|---|---|---|---|
| Name | Years | EUR/km | M-EUR/ton/hour | EUR/ton/hr |
| Pipeline [54] | 30 (15 for compressors) | 823,000 | 1.15 | 402,000 |

**Table S8: Hydrogen trucks technology assumptions**

| Technology | Lifetime | Investment Cost | OPEX | Fuel Consumption |
|---|---|---|---|---|
| Name | Years | MEUR-unit | EUR/km | H2 kg/km |
| Liquid trucks [54] | 30 | 1 | 1.4 | 0.09 |

**Table S9: Electrical Demand Data. Current loads based on [45], 2040 loads based on [28] except for UK [30].**

| **Country** | **Current Loads 2016 (TWh)** | **2040 Loads (TWh)** | **% Increase** |
|---|---|---|---|
| Germany | 508 | 785 | 55 |
| Spain | 246 | 320 | 30 |
| France | 486 | 583 | 20 |
| UK | 336 | 440 | 31 |
| Italy | 322 | 386 | 20 |

**Table S10: Hydrogen Demand Data. Base demand based on [28], except for UK [29].**

| **Country** | **Base Demand (Mtonnes/year)** | **Aviation Demand (Mtonnes/year)** |
|---|---|---|
| Germany | 4.47 | 0.71 |
| Spain | 0.82 | 0.95 |
| France | 0.92 | 0.85 |
| GB | 2.40 | 0.91 |
| Italy | 2.15 | 0.90 |
| Total | **10.76** | **4.33** |

## Supplemental Figures

**Figure S1: Inputs for energy systems model A) Electricity demand in 2040 B) Current electricity generation and transmission network (not to scale), based on year 2016**

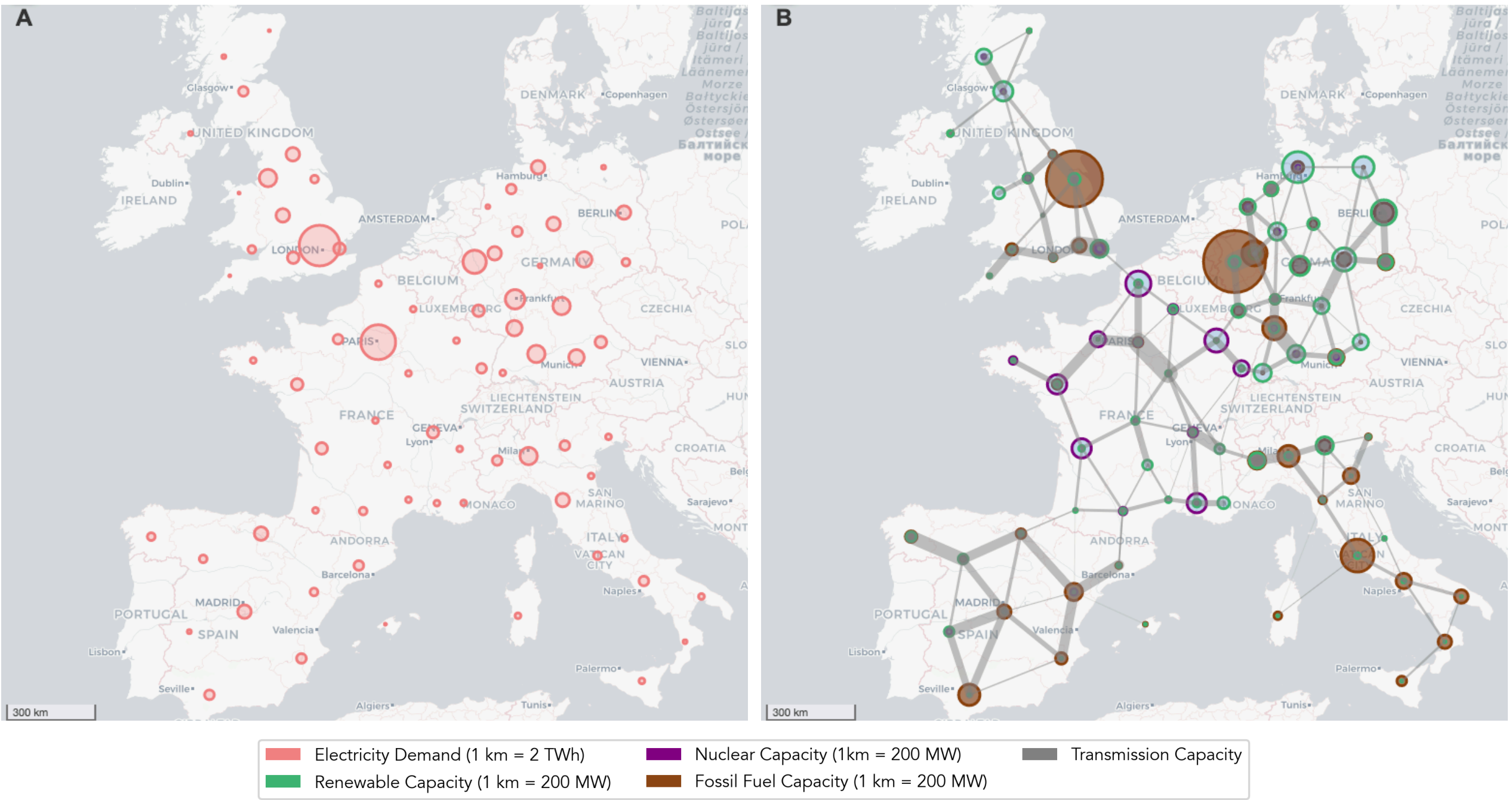

**Figure S2: Comparing different energy system scenarios. All scenarios consider Base + Aviation hydrogen demand, except S0 – Base demand only. A) Electrical capacity B) Electrical generation C) Hydrogen capacity D) Hydrogen production E) Storage capacity F) Transmission capacity**

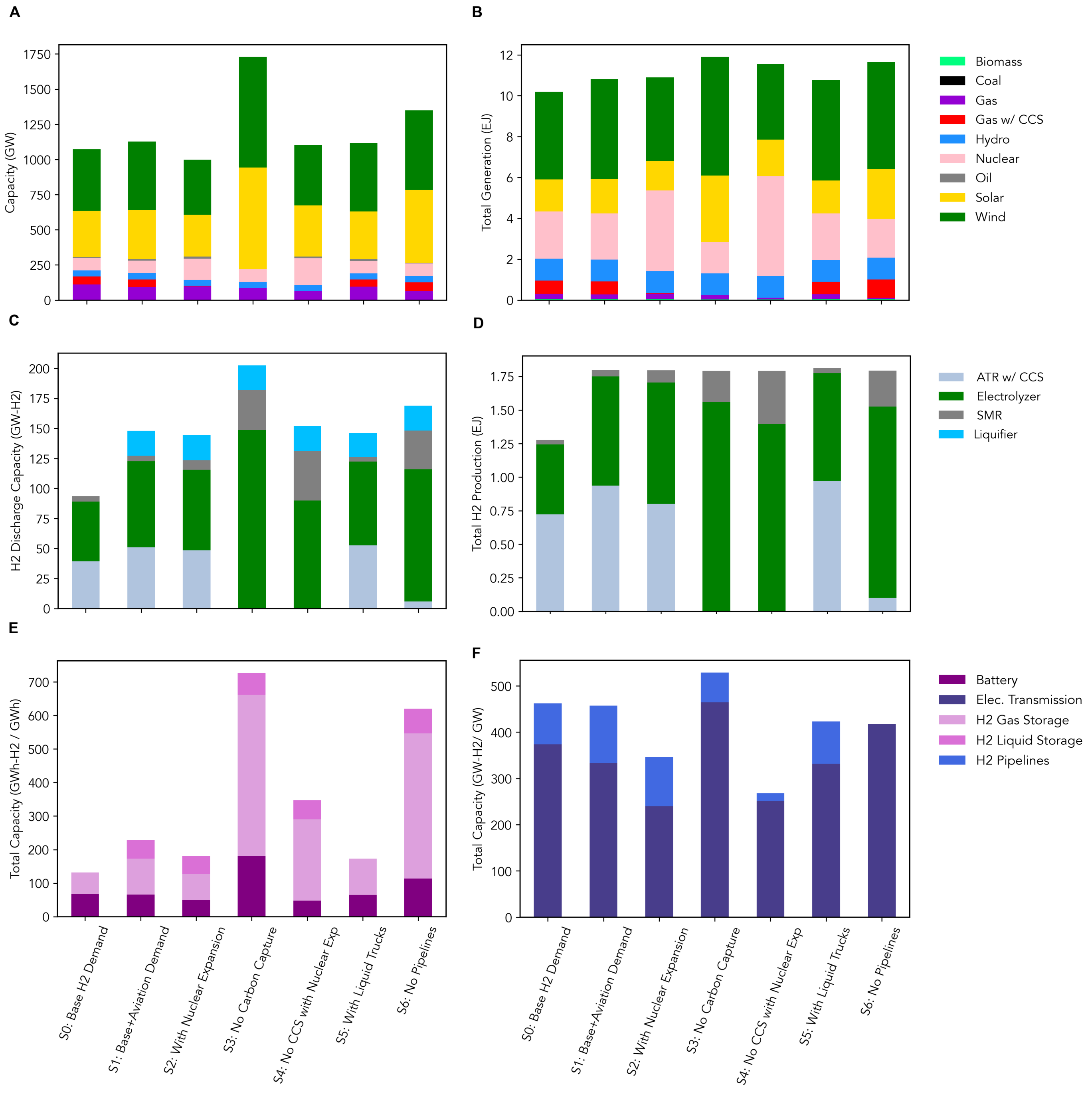

**Figure S3: Base H2 Demand scenario A) Hydrogen system map B) Electrical system map**

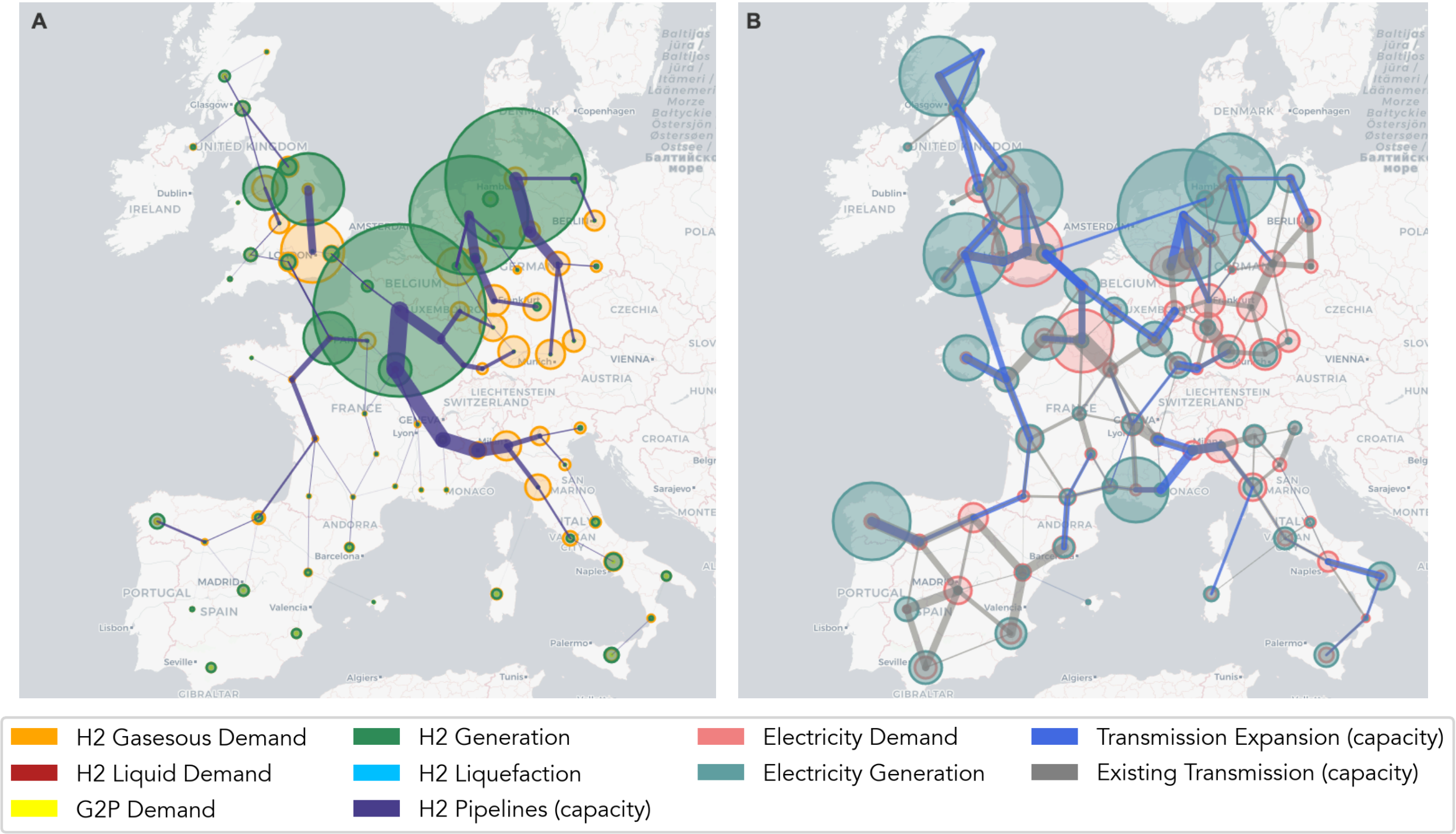

**Figure S4: No CCS with Nuclear Expansion scenario A) Hydrogen system map B) Electrical system map**

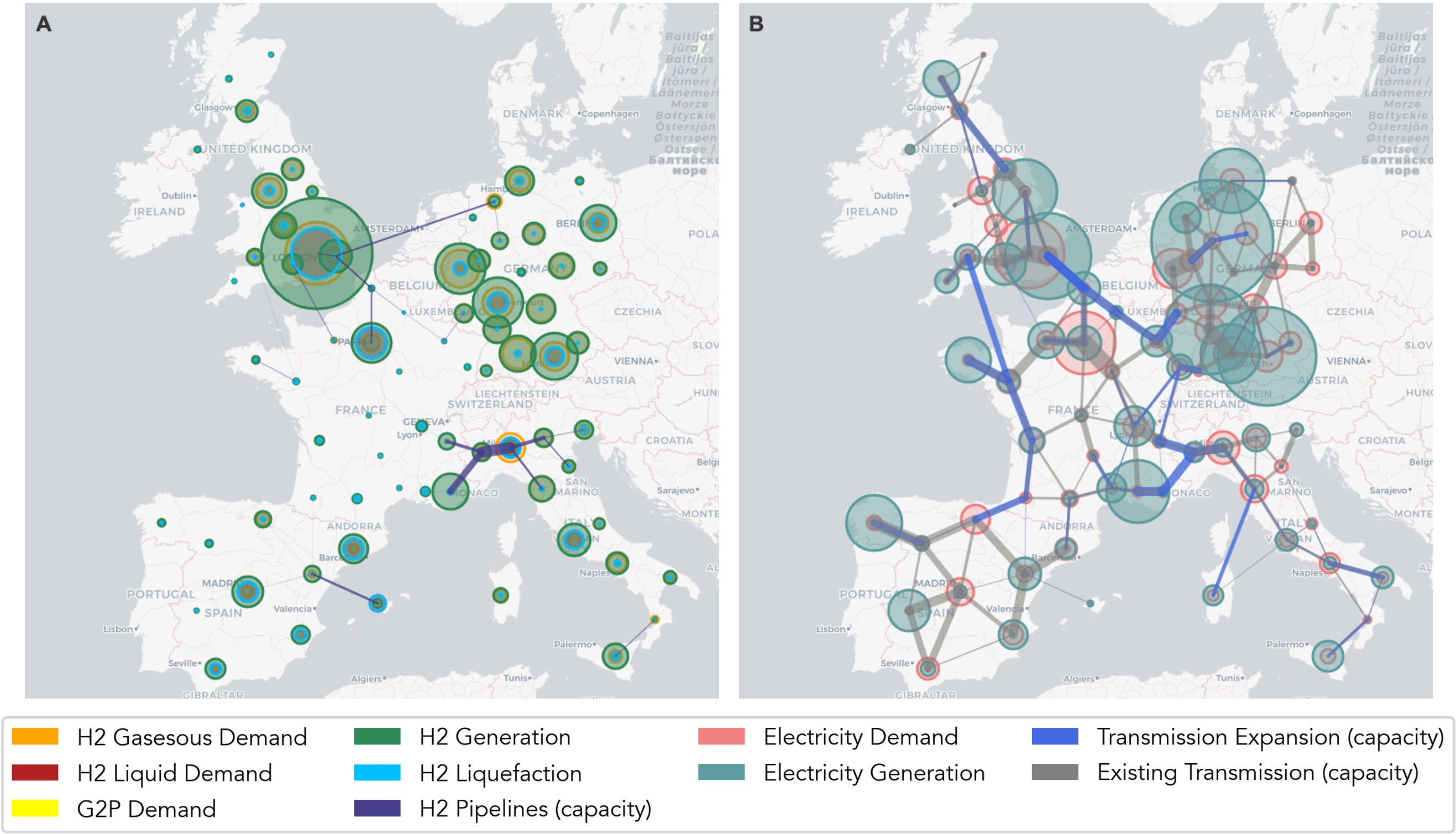

**Figure S5: Electrolytic hydrogen production, liquefaction, demand, and electricity price per zone for the No CCS with Nuclear Expansion scenario**

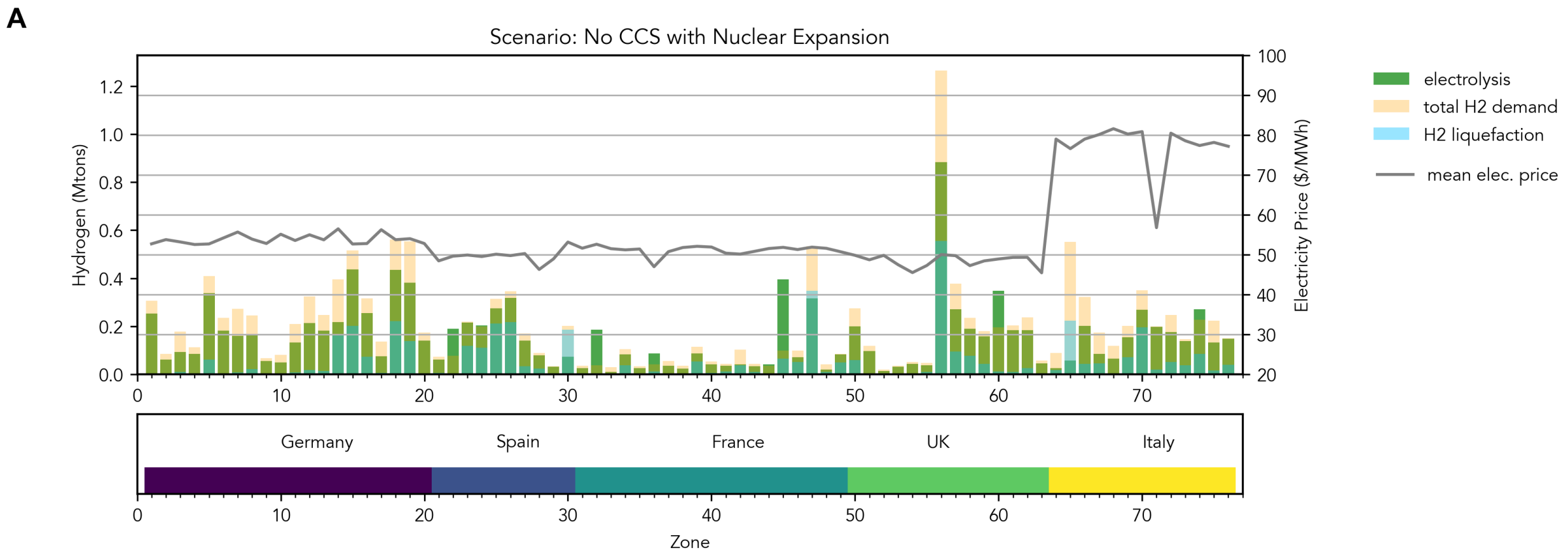

**Figure S6: A) Change in costs compared to Base + Aviation scenario B) Annual system costs C) Change in levelized cost of electricity compared to Base + Aviation scenario (system average) and D) Levelized cost of hydrogen (system average; production also includes transmission).**

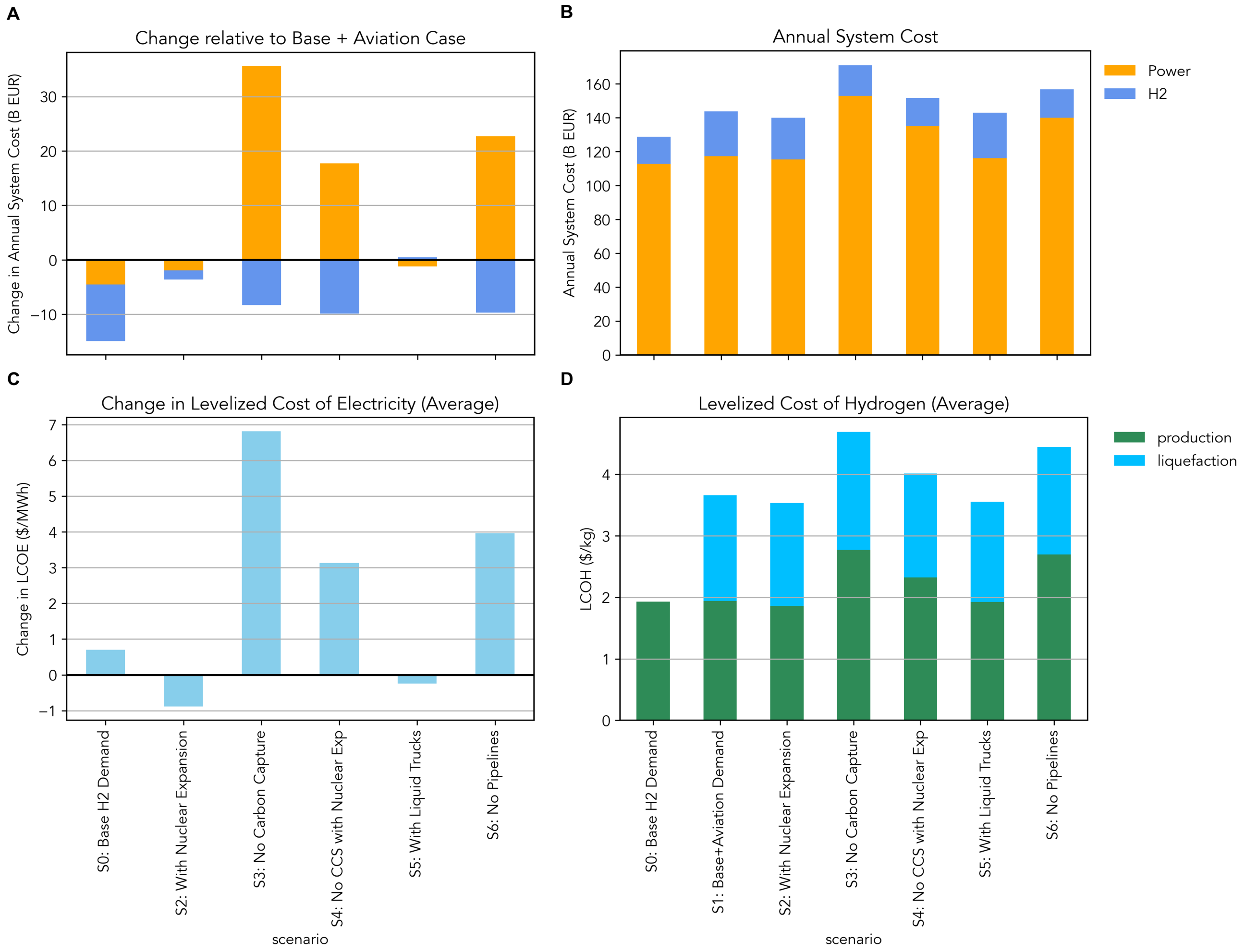

**Figure S7: Power (A) and Hydrogen (B) dispatch in the Nuclear Expansion scenario during a representative summer week. Expanding nuclear results in 130 GW less total electrical capacity than the Base + Aviation scenario. Electrolysis can be ramped up when excess renewable electricity is available, while SMR/ATR is ramped up when endogenous electrical load is high and renewables availability is limited.**

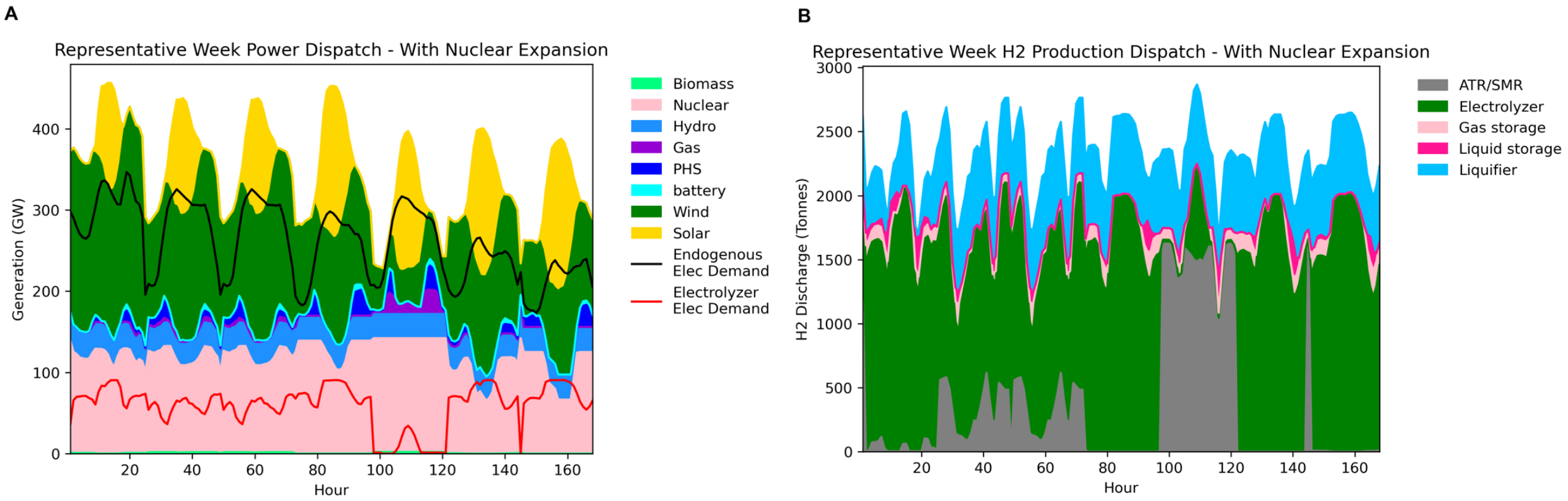